\documentclass[prl,twocolumn,superscriptaddress]{revtex4-1}

\usepackage{fancyhdr}
\usepackage{overpic}
\usepackage{verbatim}
\usepackage{graphics}
\usepackage{amsmath}
\usepackage[ansinew]{inputenc}
\usepackage{color}
\usepackage{rotating}
\newcommand{\ud}{\mathrm{d}}
\newcommand{\mr}{\mathbf{r}}

\newcommand{\mG}{\mathbf{G}}

\newcommand{\mE}{\mathbf{E}}

\newcommand{\tlo}{\tilde{\omega}}

\newcommand{\me}{\mathbf{e}}
\newcommand{\mf}{\mathbf{f}}
\newcommand{\mft}{\tilde{\mathbf{f}}}

\begin{document}

\title{Effective mode volumes and Purcell factors for leaky optical cavities}
\author{Philip Trøst Kristensen}
\affiliation{DTU Fotonik, Technical University of Denmark, DK-2800 Kgs. Lyngby, Denmark}
\affiliation{Department of Physics, Queen's University, Ontario, Canada K7L 3N6}
\author{Cole Van Vlack}
\affiliation{Department of Physics, Queen's University, Ontario, Canada K7L 3N6}
\author{Stephen Hughes}
\affiliation{Department of Physics, Queen's University, Ontario, Canada K7L 3N6}

%\author{Philip Tr{\o}st Kristensen}
%\author{Jesper M{\o}rk}
%\author{Peter Lodahl}
%\affiliation{DTU Fotonik, Technical University of Denmark, DK-2800 Kgs. Lyngby, Denmark}
%\author{Stephen Hughes}
%\affiliation{Department of Physics, Queen's University, Ontario, Canada K7L 3N6}

\date{\today}

\begin{abstract}
We show that for optical cavities with any finite dissipation, the term ``cavity mode'' should be understood as a solution to the Helmholtz equation with outgoing wave boundary conditions. This choice of boundary condition renders the problem non-Hermitian, and we demonstrate that the common definition of an effective mode volume is ambiguous and not applicable. Instead, we propose an alternative effective mode volume which can be easily evaluated based on the mode calculation methods typically applied in the literature. This corrected mode volume is directly applicable to a much wider range of physical systems, allowing one to compute the Purcell effect and other interesting optical phenomena in a rigorous and unambiguous way.
\end{abstract}

\maketitle
%\fancyhead[CO,CE]{\textcolor{red}{Confidential}}
%\thispagestyle{fancy}

%================================== Introduction ===================================%
Optical microcavities are inherently dissipative and are typically characterized by a quality factor, or $Q$-value, describing the relative energy loss per cycle as well as an effective mode volume, $V_\text{eff}$, which gives a measure of the spatial confinement of the electromagnetic field in the cavity. Cavities with high $Q$-values and small mode volumes provide enhanced light-matter interaction and are of fundamental as well as technological interest \cite{ChangCampillo1996,Nature_424_839_2003}. Effective mode volumes are ubiquitous in physics and connect to a wide range of phenomena, including sensing \cite{Loncar_APL_82_4648_2003}, switching \cite{APL_94_021111_2009}, cavity quantum electrodynamics (QED) \cite{CarmichaelBook2}, circuit-QED \cite{circuit}, and  optomechanics \cite{painter}. As a striking example of the use of mode volumes, an emitter in an optical cavity will experience a medium-enhanced radiation rate relative to that in a homogeneous medium given by the so-called Purcell factor \cite{Purcell_PR_69_681_1946}
\begin{align}
F_\text{P} = \frac{3}{4\pi^2}\left(\frac{\lambda_\text{c}}{n_\text{c}}\right)^3\left(\frac{Q}{V_\text{eff}}\right),
\label{Eq:PurcellFactor}
\end{align}
where $\lambda_\text{c}$ is the free space wavelength, and $n_\text{c}$ is the material refractive index at the field antinode $\mr_\text{c}$. Purcell factors are widely used in quantum optics as a figure of merit for single photon sources \cite{kiraz}.

\begin{figure}[htb]
\includegraphics[width=\columnwidth]{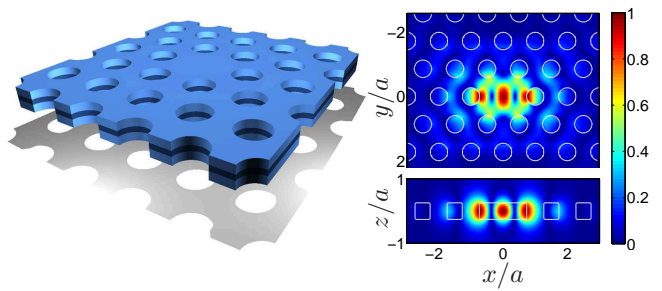}
% \begin{figure}[htb]
% \flushright
% \begin{overpic}[width=3.6cm]{L1_absField.eps}
% \put(-142,-6){\includegraphics[width=5cm]{PCslab_L1cavity_small.eps}}
% \put(38,-7){{$x/a$}}
% \put(-5,12){{\begin{sideways}{$z/a$}\end{sideways}}}
% \put(-5,54){{\begin{sideways}{$y/a$}\end{sideways}}}
% \end{overpic}\\[1mm]
\vspace{-5mm}
\caption{\label{Fig:3DsketchPlusModes}Sketch of a photonic crystal made from a triangular lattice of air holes (lattice constant $a$) in a membrane of high refractive index. A defect cavity is formed by the omission of a single hole. Right: Absolute value of the $y$-polarized
cavity mode in the planes $z=0$ (top) and $y=0$ (bottom).}
\end{figure}

Mode volumes are often attributed to the physically appealing idea of a single cavity mode. However, in spite of the fact that cavity modes are widely used in the literature, there seems to be a disturbing lack of a precise definition, and their mathematical properties therefore remain unspecified. The lack of a definition is evidenced in part by the diverse nomenclature at use (``resonance'', ``leaky mode'' or ``quasi mode''), suggesting that the dissipative nature of cavity modes somehow makes them different from other modes, but an explicit distinction is rarely made. It appears that cavity modes are widely believed to share the properties of Hermitian eigenvectors, although a mode with a finite lifetime is incompatible with the solution space of Hermitian eigenvalue problems. Strictly, only in the limit of large (infinite) $Q$ will the modes in optical cavities appear as the solutions to Hermitian eigenvalue problems. For finite $Q$, the lack of hermiticity effectively renders expressions such as Eq.~(\ref{Eq:PurcellFactor}) ambiguous, since the volume $V_\text{eff}$ cannot be inferred from the usual inner product of Hermitian systems. In particular, if $\epsilon_\text{r}(\mr)$ describes the relative permittivity distribution and $\mft_\text{c}(\mr)$ is the cavity mode, then a direct application of the common normal mode prescription
\begin{align}
V^\text{N}_\text{eff}=\int_V \frac{\epsilon_\text{r}(\mr) |\mf_\text{c}(\mr)|^2}{\epsilon_\text{r}(\mr_\text{c})|\mf_\text{c}(\mr_\text{c})|^2}  \ud\mr,
\label{Eq:VeffN}
\end{align}
cannot be expected to provide the correct mode volume. In many practical calculations, the leaky cavity mode is found with the finite-difference time-domain (FDTD) method by launching a short pulse and monitoring the resonant field that leaks from the cavity at a rate set by the $Q$-value \cite{Yao_LaserAndPhotonicsReviews_2009}. Figure \ref{Fig:3DsketchPlusModes} shows a sketch of an example cavity along with mode profiles calculated with FDTD \cite{lumerical}. For this cavity mode the integral in Eq. (\ref{Eq:VeffN}) diverges as a function of the integration volume $V$, and indeed this is formally the case for all cavities with a finite $Q$. For very high-$Q$ cavities, however, the divergence is slow and may not be discernable in practice due to numerical accuracy, but the formal divergence still renders Eq. (\ref{Eq:VeffN}) questionable.

In this Letter we argue that the term ``cavity mode'' should be understood as a solution to the Helmholtz equation with outgoing wave boundary conditions. This definition renders the cavity modes identical to the quasinormal modes of Lee \emph{et al.}~\cite{Lee_JOSAB_16_1409_1999} which have complex resonance frequencies (with a negative imaginary part, as expected for dissipative modes) and exhibit an inherent exponential divergence at large distances. We illustrate directly how this definition complies with typical calculations of cavity modes using FDTD and we elucidate how the cavity modes from FDTD calculations also show an exponential divergence. Quasinormal modes have properties that are different from the solutions to Hermitian eigenvalue problems and therefore important results such as orthogonality and completeness of the solutions cannot be taken for granted. Nevertheless, an inner product and a corresponding orthogonality relation can be defined, and the quasinormal modes can be used as a basis for expansion of the electromagnetic Green's tensor in certain regions \cite{Lee_JOSAB_16_1409_1999}. This enables a precise and unambiguous description of light-matter interaction in general leaky optical cavities, including an expression for the effective mode volume and thus the Purcell factor.

%================================== Definition of modes ===================================%
It is illustrative to start by highlighting the differences between two types of modes that can be associated with fields in optical cavities. The electric field in general non-magnetic materials satisfy the wave equation
%\begin{equation}
%\nabla\times\nabla\times\mE(\mr,t) + \frac{\epsilon_\text{r}(\mr)}{c^2}\frac{\partial}{\partial t^2}\mE(\mr,t)  %= 0,
%\label{Eq:maxwellsEquationTimeDomain}
%\end{equation}
with time-harmonic solutions of the form $\mE(\mr,t) = \mE(\mr,\omega)\exp\{-i\omega t\}.$ The position-dependent electric field $\mE(\mr)$ solves the vector Helmholtz equation
\begin{equation}
\nabla\times\nabla\times\mE(\mr,\omega) - k_0^2\epsilon_\text{r}(\mr)\mE(\mr,\omega) = 0,
\label{Eq:Helmholtz}
\end{equation}
where $k_0=\omega/c$. Together with a suitable set of boundary conditions, Eq. (\ref{Eq:Helmholtz}) provides a generalized eigenvalue equation. We will use the term {\em normal mode} to denote a solution to Eq.~(\ref{Eq:Helmholtz}) with any set of boundary conditions that renders the problem Hermitian. In this case we denote the vector eigenfunctions and corresponding real eigenfrequencies as $\mathbf{f}_\mu(\mr)$ and $\omega_\mu$, respectively. The normal modes are normalized as
\begin{align}
\langle\mf_\mu|\mf_\lambda\rangle = \int_V\epsilon_\text{r}(\mr)\, \mf^*_\mu(\mr)\cdot\mf_\lambda(\mr)\,\ud\mr = \delta_{\mu,\lambda},
\label{Eq:normalizationOfNormalModes}
\end{align}
where the integral is over the volume defined by the boundary conditions. In many applications, the limit $V\rightarrow\infty$ is taken, in which case the spectrum of eigenvalues becomes continuous.
%================================== Quasinormal modes ===================================%
We will use the term {\em quasinormal modes}
%(also sometimes referred to as morphology-dependent resonances \cite{Lee_JOSAB_16_1409_1999})
for solutions to Eq.~(\ref{Eq:Helmholtz}) with outgoing wave boundary conditions (the Sommerfeld radiation condition  \cite{Martin_MultipleScattering}). This choice of boundary condition renders the eigenvalue problem non-Hermitian with a discrete spectrum, and we denote the vector eigenfunctions with a tilde as $\mft_\mu(\mr)$. The corresponding eigenfrequencies, $\tlo_\mu=\tlo_\mu^\text{R}+i\tlo_\mu^\text{I}$, are in general complex with $\tlo_\mu^\text{I}<0$, and it follows from Eq.~(\ref{Eq:Helmholtz}) that, contrary to the Hermitian case, $\mft_\mu(\mr)$ and $\mft^*_\mu(\mr)$ are
%in general
\emph{not} eigenvectors corresponding to the same eigenvalue.
The quasinormal modes may be normalized as \cite{Lee_JOSAB_16_1409_1999}
\begin{align}
\langle\langle \mft_\mu|\mft_\lambda \rangle\rangle = &\lim_{V\rightarrow\infty}\int_{V} \epsilon_\text{r}(\mr)\, \mft_\mu(\mr)\cdot \mft_\lambda(\mr)\,\ud\mr  \nonumber \\
&+ i\frac{\sqrt{\epsilon_\text{r}}}{\tlo_\mu+\tlo_\lambda} \int_{\partial V}\mft_\mu(\mr)\cdot \mft_\lambda(\mr)\,\ud\mr = \delta_{\mu,\lambda},
\label{Eq:NormalizationOfQuasiModes}
\end{align}
where $\partial V$ denotes the border of the volume $V$. The limit $V\rightarrow\infty$ is calculated by increasing the volume to obtain convergence. For the systems that we investigate in this Letter, the convergence is remarkably fast. For very low-$Q$ cavities, however, the convergence is nontrivial due to the exponential divergence of the quasinormal modes which may cause the inner product to oscillate around the proper value as a function of calculation domain size.

In addition to the modes of the cavity it is convenient to introduce the electromagnetic Green's tensor through
\begin{equation}
\nabla\times\nabla\times\mG(\mr,\mr',\omega) - k_0^2\epsilon_\text{r}(\mr)\mG(\mr,\mr',\omega) = \mathbf{I}\delta(\mr-\mr'),
\label{Eq:GWaveEquation}
\end{equation}
subject to the Sommerfeld radiation condition. The Green's tensor provides the proper framework for calculating light emission and scattering in general dielectric structures. In general, the decay rate $\Gamma_\alpha(\mr,\omega)$ of a dipole emitter with orientation $\me_\alpha$ may be enhanced or suppressed as compared to the rate $\Gamma_\text{B}$ in a homogeneous medium. The relative rate may be expressed as
\begin{align}
\frac{\Gamma_\alpha(\mr,\omega)}{\Gamma_\text{B}(\omega)} = \frac{\text{Im}\left\{\me_\alpha\mG(\mr,\mr,\omega)\me_\alpha\right\}}{\text{Im}\left\{\me_\alpha\mG_\text{B}(\mr,\mr,\omega)\me_\alpha\right\}},
\label{Eq:PurcellFromLDOS}
\end{align}
where $\mG_\text{B}(\mr,\mr',\omega)$ is the Green's tensor in a homogeneous background medium with $\epsilon_\text{r}(\mr)=\epsilon_\text{B}$ \cite{Martin_PRE_58_3909_1998}. In certain regions, such as inside the scattering region, the transverse part of the Green's tensor may be expanded
through \cite{Lee_JOSAB_16_1409_1999}
\begin{align}
\mG^\text{T}(\mr,\mr',\omega) = c^2\sum_\mu\frac{\mft_\mu(\mr)\mft_\mu(\mr')}{2\tlo_\mu(\tlo_\mu-\omega)}.
%\quad (\mr,\mr') \in \mathcal{D}_\text{Q}^2.
\label{Eq:GreensTensorFromQuasiNormalModes}
\end{align}
The implicit assumption behind the notion of a cavity mode is that one term $\mu=\text{c}$ dominates the expansion in Eq.~(\ref{Eq:GreensTensorFromQuasiNormalModes}) and hence that the Green's tensor can be approximated by this term only. With this assumption, and noting that $\text{Im}\{\mG(\mr,\mr,\omega)\} = \text{Im}\{\mG^\text{T}(\mr,\mr,\omega) \}$, one can use Eqs.~(\ref{Eq:PurcellFromLDOS}) and (\ref{Eq:GreensTensorFromQuasiNormalModes}) with $\omega=\omega_\text{c}=\tlo_\text{c}^\text{R}$ to recover Eq.~(\ref{Eq:PurcellFactor}) with a \emph{corrected} effective mode volume given as
\begin{align}
V_\text{eff}^\text{Q} = \frac{1}{n_\text{c}^2}\frac{|v_\text{Q}|^2}{v_\text{Q}^\text{R}},\quad v_\text{Q}=\frac{\langle\langle\mft_\text{c}|\mft_\text{c}\rangle\rangle}{\mft^2_\text{c}(\mr_\text{c})},
\label{Eq:VeffQ}
\end{align}
where $v_\text{Q}=v_\text{Q}^\text{R}+iv_\text{Q}^\text{I}$ is complex in general. This prescription provides a direct and unambiguous way of calculating the effective mode volume.

The quasinormal modes can be calculated analytically for sufficiently simple structures, but for general structures the outgoing wave boundary conditions are not immediately compatible with standard numerical solution methods. One option is to rewrite Eq. (\ref{Eq:Helmholtz}) as
\begin{align}
\nabla\times\nabla\times\mE(\mr,\omega) - k_0^2\epsilon_\text{B}\mE(\mr,\omega) = k_0^2\Delta\epsilon(\mr)\mE(\mr,\omega),
\end{align}
%$
%\nabla\times\nabla\times\mE(\mr) - k_0^2\epsilon_\text{B}(\mr)\mE(\mr) = k_0^2\Delta\epsilon(\mr)\mE(\mr),
%$
where $\Delta\epsilon=\epsilon_\text{r}(\mr)-\epsilon_\text{B}$, and calculate the quasinormal modes
from
%as solutions to
a Fredholm type integral equation,
\begin{align}
\mE(\mr,\omega)=\left(\frac{\omega}{c}\right)^2\int_V\mG^\text{B}(\mr,\mr',\omega)\,\Delta\varepsilon(\mr')\,\mE(\mr',\omega)\ud \mr',
\label{Eq:LippmannSchwingerEigenvalue}
\end{align}
%Although Eq. (\ref{Eq:LippmannSchwingerEigenvalue}) is complicated by the nontrivial dependence of the integral operator on the eigenvalue, it
which manifestly respects the outgoing wave boundary conditions. Another option is to use FDTD to calculate the quasinormal mode as the resonant field that is excited by an initial short input pulse. We have used both Eq.~(\ref{Eq:LippmannSchwingerEigenvalue}) and FDTD to calculate the quasinormal modes in different example cavities in two and three dimensions.
%The 2D examples serve to illustrate the method and allows for a direct comparison between the two calculation methods, while the
%3D calculation provides a direct
% application example.
 For our example below,
 %For the results in this Letter,
 we  solve Eq.~(\ref{Eq:LippmannSchwingerEigenvalue}) using the expansion technique of Ref.~\cite{Kristensen_JOSAB_27_228_2010} with an additional iteration of $k_0$ to make the solution self-consistent.
% no room just now
%To this end, Eq. (\ref{Eq:LippmannSchwingerEigenvalue}) is recast as a matrix eigenvalue equation of the form $\mG(\lambda_{n-1})\me_n$=$\lambda_n\me_n$, with $\lambda$=$1/k_0^2$. In each iteration step, the matrix $\mG(\lambda_{n-1})$ is constructed based on the solution in the previous step, and the iteration continues until a self-consistent solution is achieved. The
 In addition, we perform FDTD calculations using perfectly matched layers (PMLs) \cite{Taflove_1995} to enforce the outgoing wave bounday conditions.
  %and a run-time fast Fourier transform \cite{Kim_PRB_73_235117_2006} to reveal the mode profile.
 % at the resonance frequency.

\begin{figure}[htb]
\includegraphics[width=\columnwidth]{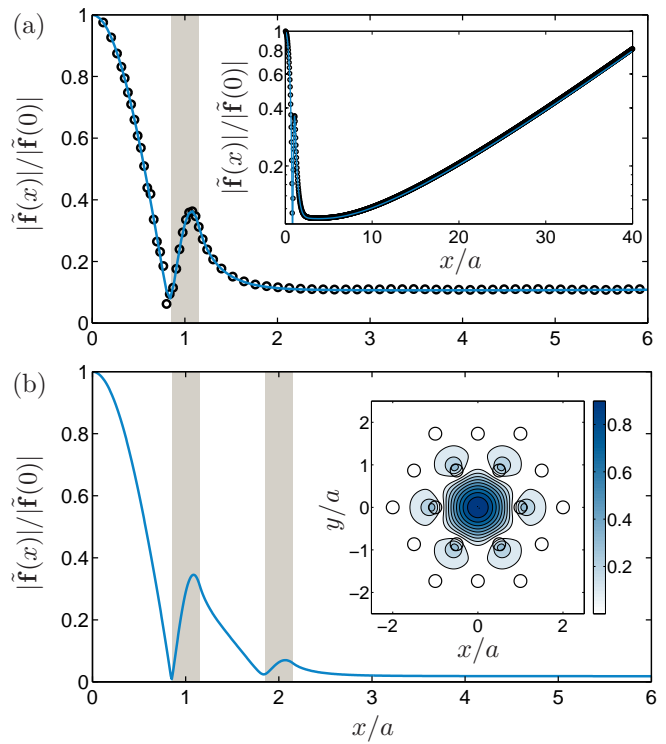}
% \flushright
% \begin{overpic}[width=8cm]{cavityMode_1layer_alongXaxis_wFDTD_v3.eps}
% \put(-8,52){(a)}
% \put(-8,19){\begin{sideways}{$|\mft(x)|/|\mft(0)|$}\end{sideways}}
% \put(33,17){\includegraphics[width=5.1cm]{L1_fdtdCompare_fix.eps}}
% \put(63,13){{$x/a$}}
% \put(26,25){{\begin{sideways}{$|\mft(x)|/|\mft(0)|$}\end{sideways}}}
% \end{overpic}\\[1mm]
% \begin{overpic}[width=8cm]{cavityMode_2layers_alongXaxis_v2.eps}
% \put(-8,52){(b)}
% \put(48.5,-5){$x/a$}
% \put(-8,19){\begin{sideways}{$|\mft(x)|/|\mft(0)|$}\end{sideways}}
% \put(48,12){\includegraphics[width=3.8cm]{cavityMode_2layers_forInset.eps}}
% \put(66.5,8){{$x/a$}}
% \put(43,29.5){{\begin{sideways}{$y/a$}\end{sideways}}}
% \end{overpic}\\[2mm]
\vspace{-5mm}
\caption{\label{Fig:cavityMode_2layers}(a): Absolute value along the $x$-axis of the quasinormal mode in the 2D crystallite for the case of $N$=1. Blue solid line shows the solution to Eq. (\ref{Eq:LippmannSchwingerEigenvalue}), and black circles show the calculation using FDTD. Inset shows long distance behavior on a logarithmic scale. (b): Absolute value along the $x$-axis of the quasinormal mode for the case of $N=2$ with the inset showing the distribution in the $xy$-plane. Grey shaded areas indicate the high-index rods.}
\end{figure}

\begin{figure}[htb]
\includegraphics[width=\columnwidth]{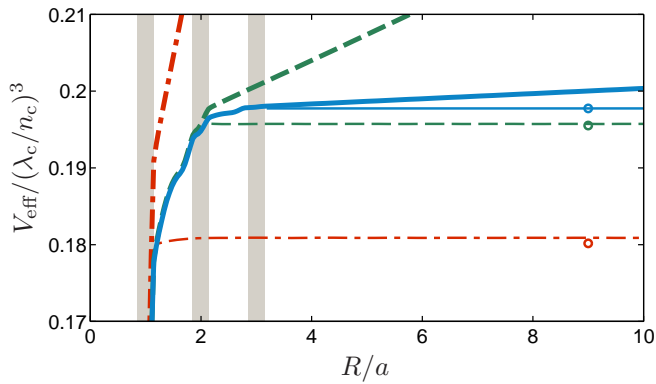}
% \begin{figure}
% \flushright
% \begin{overpic}[width=8cm]{effectiveModeVolume_1and2and3Layers_fixed.eps}%effectiveModeVolume_2and3Layers_corrected.eps}
% \put(48,-5){$R/a$}
% \put(-6,18){\begin{sideways}{$V_\text{eff}/(\lambda_\text{c}/n_\text{c})^3$}\end{sideways}}
% %\put(-6,23){\begin{sideways}{$V_\text{eff}/\lambda_\text{c}^2$}\end{sideways}}
% \end{overpic}\\[2mm]
\vspace{-5mm}
\caption{\label{Fig:modeVolume_2Layers}Effective mode volumes $V_{\rm eff}^\text{N}$ (thick lines) and $V_{\rm eff}^\text{Q}$ (thin lines) for $N=1$ (red dash-dotted), $N=2$ (green dashed) and $N=3$ (blue solid) as a function of radius $R$ of the calculation domain. Circles indicate reference mode volumes $V_\text{eff}^\text{tot}$ from independent Green's tensor calculations \cite{Kristensen_JOSAB_27_228_2010}, and grey dashed areas show the rod cross sections along the $x$-axis.}
\end{figure}

%================================== Example calculation ===================================%
We first consider a 2D finite-sized hexagonal crystallite of high-index rods in air with a single missing rod in the center. The rods have relative permittivity  $\epsilon_\text{r}$=$11.4$ and radius $R$=$0.15\,a$, where $a$ is the lattice constant. We focus on transverse magnetic (TM) polarization in which the electric field is in the direction of the rods. In the limit of infinite size, the photonic crystal exhibits a photonic band gap \cite{Joannopoulos2008}, and the $Q$ of the cavity therefore depends on the size of the structure which we may characterize by the number of rod layers, $N$. For $N$=1 and $N$=2, Fig.~\ref{Fig:cavityMode_2layers} shows the supported cavity modes with frequencies $\tlo_\text{c}a/2\pi c$=$0.4259-0.0135i$ ($Q=-\tlo_\text{R}/2\tlo_\text{I}\approx16$) and $\tlo_\text{c}a/2\pi c=0.4218-0.0013i$ ($Q\approx163$). In the case of $N$=3 (not shown), the structure supports a cavity mode with frequency $\tlo_\text{c}a/2\pi c=0.4216-0.0001i$ ($Q\approx1576$).
As expected, the quasinormal modes are concentrated in the center of the cavity and seem to fall off with increasing distance to the crystallite. At large distances, however, the quasinormal modes (by definition) behave as outgoing waves of the form $\mft(\mr)\propto\exp(ik_0r)/\sqrt{r}$ (2D) and $\mft(\mr)\propto\exp(ik_0r)/r$  (3D), and since $k_0=k_\text{R}+ik_\text{I}$ with $k_\text{I}<0$, they diverge exponentially as $r\rightarrow\infty$. For the case of $N$=1, the top panel in Fig.~\ref{Fig:cavityMode_2layers} illustrates directly that the solutions to Eq.~(\ref{Eq:LippmannSchwingerEigenvalue}) are identical to those obtained from FDTD. In particular, both solution methods pick up the divergence in the field at large distances.
%================================== Mode volumes ===================================%
Figure \ref{Fig:modeVolume_2Layers} shows, as a function of the size of the calculation domain, the corrected effective mode volume in Eq.~(\ref{Eq:VeffQ}) along with the common definition in Eq.~(\ref{Eq:VeffN}) with $\mf_\text{c}(\mr)=\mft_\text{c}(\mr)$. From the figure it is clear that whereas $V_\text{eff}^\text{Q}$ converges quickly to the limiting values, $V_\text{eff}^\text{N}$ seems to increase with the size of the domain.
 %$V_{\rm eff}^\text{Q}/\lambda_\text{c}^2=0.196$ ($L=2$) and $V_{\rm eff}^\text{Q}/\lambda_\text{c}^2=0.198$ ($L=3$). For TM polarization in 2D the Purcell factor is given as $F_\text{P}=\lambda_\text{c}^2Q/n_\text{c}^2\pi^2V_\text{eff}$, and the resulting Purcell factors are 84.21 {\bf *probably one decimal place is enough, and maybe these can also be shown on th graph, perhaps at the right of the axes*} and 811.5, respectively. These values should be compared to the numerically exact values 84.20 and 807.6 from direct Green's tensor calculations \cite{Kristensen_JOSAB_27_228_2010}.
The linear divergence in $V_{\rm eff}^\text{N}$ with the size of the normalization domain was also noted in Ref. \cite{Koenderink_OL_35_4208_2010} and derives from the fact that the field does not go to zero at positions outside the crystallite, cf.~Fig.~\ref{Fig:cavityMode_2layers}. At much larger $V$, the field, and hence $V_{\rm eff}^\text{N}$, diverges exponentially. For increasing Q, the linear divergence with domain size becomes less and less pronounced, suggesting that the two formalisms converge in the limit of infinite $Q$ as expected. 

\begin{figure}[htb]
\includegraphics[width=\columnwidth]{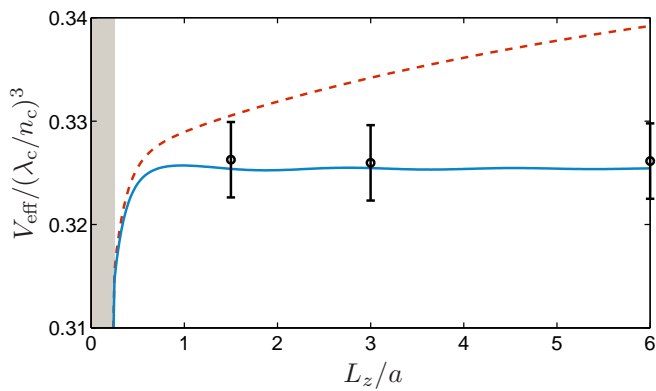}
% \begin{figure}[htb]
% \flushright
% \begin{overpic}[width=8cm]{L1_Veff_error_new.eps}
% \put(48,-5){$L_z/a$}
% \put(-6,18){\begin{sideways}{$V_\text{eff}/(\lambda_\text{c}/n_\text{c})^3$}\end{sideways}}
% \end{overpic}\\[2mm]
\vspace{-5mm}
\caption{\label{Fig:modeVolume_3D}Effective mode volume $V_{\rm eff}^\text{N}$ (red dashed) and $V_{\rm eff}^\text{Q}$ (blue solid) for the cavity in Fig. \ref{Fig:3DsketchPlusModes} as a function of height of the calculation domain. Circles indicate reference mode volumes $V_\text{eff}^\text{tot}$ derived from independent Green's tensor calculations \cite{Yao_LaserAndPhotonicsReviews_2009} with estimated error bars at different domain heights. Gray dashed area shows the extend of the membrane.}
\end{figure}

Next, for a practical 3D example we consider a photonic crystal membrane ($\epsilon_\text{r}$=$12$) of thickness $h=0.5\,a$ and hole radius $r=0.275a$. A single air hole is omitted to create a cavity, and Fig. \ref{Fig:3DsketchPlusModes} shows the supported cavity mode with frequency $\tlo a/2\pi c$=$0.2904-0.0004i$ ($Q\approx362$). From the results in two dimensions we know that the quasinormal modes can be directly calculated using FDTD with PMLs. The field in Fig. \ref{Fig:3DsketchPlusModes} was calculated in the same way, and we therefore argue that it is indeed a quasinormal mode and that we should use Eq.~(\ref{Eq:VeffQ}) rather than Eq.~(\ref{Eq:VeffN}) to calculate the effective mode volume. Fig. \ref{Fig:modeVolume_3D} shows both $V_{\rm eff}^\text{Q}$ and $V_{\rm eff}^\text{N}$ as a function of calculation domain size. At the quasinormal mode frequency, the photonic band gap prevents in-plane propagation, and therefore the only way for the field to leak out of the cavity is in the $z$-direction. This means that both $V_{\rm eff}^\text{Q}$ and $V_{\rm eff}^\text{N}$ converge quickly as a function of width and depth of the calculation domain and we focus only on the variation in the effective mode volumes with the height of the calculation domain. As in two dimensions, the data shows a fast convergence of $V_{\rm eff}^\text{Q}$, while $V_{\rm eff}^\text{N}$ clearly diverges, confirming that Eq.~(\ref{Eq:VeffN}) is not applicable.

Finally, we compare the calculated mode volumes to independent calculations using the Green's tensor \cite{Kristensen_JOSAB_27_228_2010,Yao_LaserAndPhotonicsReviews_2009}. Substituting $F_\text{P}=\Gamma_\text{c}(\mr,\omega_\text{c})/\Gamma_\text{B}(\omega_\text{c})$ in the expression for the Purcell factor \cite{note1} defines an effective mode volume $V_\text{eff}^\text{tot}$. For each of the cavities, $V_\text{eff}^\text{tot}$ is indicated with a circle in Fig.~\ref{Fig:modeVolume_2Layers}. The maximum estimated absolute error in these calculations is less than 0.0003. The observable discrepancies for $N$=1 and $N$=2 stem from the single mode approximation and indicates the limited validity of the Purcell factor. In Fig.~\ref{Fig:modeVolume_3D}, $V_\text{eff}^\text{tot}$ was calculated with FDTD as the response to an input dipole source at three different domain sizes and with estimated error bars as indicated. These independent calculations confirm that Eq.~(\ref{Eq:VeffQ}) not only is unambiguous, but also leads to the correct value within the single mode approximation.

%================================== Conclusion ===================================%
In conclusion, we have shown that the term ``cavity mode'' should be understood as a so-called quasinormal mode, defined as a solution to the Helmholtz equation with outgoing wave boundary conditions. This can have profound consequences, since this choice of boundary conditions renders the differential equation problem non-Hermitian so that common results from Hermitian eigenvalue analysis do not apply. In particular, the quasinormal modes have complex frequencies and exhibit an inherent divergence at long distances which makes the calculation of an effective mode volume nontrivial. Introducing an inner product that carefully accounts for the long distance behavior, it is possible to normalize the quasinormal modes and to define an effective mode volume in a direct and unambiguous way. In practical calculations, this corrected mode volume can be obtained in a straightforward way using exactly the same cavity modes that are typically computed for use in mode volume calculations.

% This definition is slightly different from the mode volume for Hermitian modes that is commonly used in the literature, but is just as easy to evaluate and is applicable to a much wider range of physical systems. We have explicitly compared the two different mode volumes for a number of example material problems in two and three dimensions.
 %As the $Q$-value of the cavities are increased, our results suggest that the effective mode volumes from the two different methods converge to the same value as expected.

This work was supported by NSERC and The Danish Council for Independent Research (FTP 10-093651).
%We also thank {\bf *TBD*}

%==================================BACK MATTER STUFF===================================%
%\bibliography{bib1}

\begin{thebibliography}{10}
\bibitem{note1}%
  \BibitemOpen
  \bibinfo {note} {For TM polarization in 2D, the Purcell factor is given as
  $F_\text{P}=\lambda_\text{c}^2Q/(n_\text{c}^2\pi^2V_\text{eff})$. In 3D we
  use Eq. (\ref{Eq:PurcellFactor}).}%
  \bibAnnoteFile{Stop}{note1}%
\end{thebibliography}

%Merlin.mbs v4.21 2009-07-09.
%

\end{document}